\documentclass[runningheads]{llncs}
\usepackage[T1]{fontenc}
\usepackage{graphicx}
\usepackage{tabularx}
\usepackage{caption}
\usepackage{subcaption}
\usepackage{tabularx}
\usepackage[table,xcdraw]{xcolor}
\usepackage{array}
\usepackage{threeparttable}
\usepackage{url}
\begin{document}
\title{Women have it Worse: an ICT Workplace \\ Digital Transformation Stress Gender Gap}
\titlerunning{Women Have it Worse: an ICT Workplace DTS Gender Gap}
%
\author{Ewa Makowska-T{\l}omak\inst{1}\orcidID{0000-0002-6010-4210} \and \\
Sylwia Bedy{\'n}ska\inst{1}\orcidID{0000-0001-8255-1946} \and \\
Kinga Skorupska\inst{2}\orcidID{0000-0002-9005-0348} \and \\
Rados{\l}aw Nielek\inst{3}\orcidID{0000-0002-5794-7532}}
\authorrunning{E. Makowska-T{\l}omak et al.}
%
\institute{SWPS University of Social Sciences and Humanities, Warsaw, Poland
\email{emakowska-tlomak@swps.edu.pl}\\
\and
Polish-Japanese Academy of Information Technology, Warsaw, Poland
\and
 NASK-National Research Center, Warsaw, Poland}

\maketitle              
\begin{abstract}
Although information and communication technologies (ICT) solutions have positive outcomes for both companies and employees, the digital transformation (DT) could have an impact on the well-being of employees. The jobs of the employees became more demanding and they were expected to learn ICT skills and cope with ICT workloads and hassles. Due to negative stereotypes about women's deficiency in technology, these ICT problems could affect female and male employees differently. Thus, we predicted that this additional pressure may manifest itself in higher levels of digital transformation stress (DTS) in female employees. The results confirmed this prediction and indicated the existence of a gender gap in DTS, measured two-fold - in sentiment analysis of help desk tickets and self-report using a psychological scale. Based on these results, we explore the need to discuss possible solutions and tools to support women in ICT-heavy workplace contexts

\end{abstract}
\section{Introduction}

Digital transformation (DT) has the potential to alter the nature of production, work and increase productivity and growth, but it also changes the nature of work \cite{brussevich2018gender}. It encompasses digitizing business processes and introducing document management systems (DMS). Digital transformation is often related to organizational changes \cite{verina2019digital,zeike2019managers,hanelt2021systematic}. It requires instilling a culture that supports change while enabling the overall strategy of the company \cite{mergel2019defining,verina2019digital}. Although digital transformation delivers unquestionable advantages and benefits \cite{casalino2019digital,nambisan2019digital}, the DT process can be challenging both for organizations and their employees \cite{henriette2016digital,trenerry2021preparing}. Digital transformation in organizations is usually implemented as part of long-term projects \cite{barthel2020exploring} related to organizational and strategic changes \cite{henriette2016digital,hanelt2021systematic}. They may be under time pressure \cite{mullan2019have,zeike2019managers} and may cause increased ICT troubles, as well as employee work overload \cite{schwarzmuller2018does}. These difficulties bring us to the core of digital transformation stress \cite{makowska2022blended,makowska2023measuring}.

In our present study, we ask the question if the prevalence and severity of this type of stress may depend on gender, and if so, what are the underlying causes of such differences and how to address them. In the literature, there are findings that women, more frequently than men, experienced stress related to technology \cite{brussevich2018gender,marchiori2019individual,la2020technostress}, named technostress \cite{tarafdar2007understanding}. 
According to Eurostat data, the level of ICT skills among women and men aged 25 to 55 is very similar and shows no significant differences. More than 50 \% of both women and men have at least basic ICT competence, and more than 20 \% of the population has higher competence. Nevertheless, there is a very large gender gap in the percentage of ICT professionals.
Therefore, in our study, we examine whether women are also more exposed to digital transformation stress and aim to identify the main DTS factors affecting women at work. Because stress and DTS correlate negatively and highly with self-efficacy \cite{bandura1989human,preacher2008asymptotic,barbaranelli2017psychometric}), we focused on the identification of lack of self-efficacy among female employees in relation to ICT issues, during the DT project. We defined this lack of self-efficacy towards ICT as ICT helplessness. Then we examine whether there are significant differences in sentiment analysis, in context to ICT helplessness, between female and male employees.

 \section{Related Work}
 
 \subsection{Digital Transformation Stress of Employees}

Our review of the literature on stress, technostress \cite{tarafdar2007understanding,tarafdar2015technostress} and DT, and the research in the DT field, the digital transformation stress (DTS) \cite{makowska2021evaluating,makowska2023measuring} was identified as the employees’ reaction related directly to the specific digital transformation process itself in the workplace, related to the mode of DT management and DT introduction in the workplace as well as ICT hassles and DT project workload \cite{blazejewski2018digitalization,makowska2022blended,makowska2021evaluating}. 
The COVID-19 pandemic has accelerated digital transformation in nearly all sectors (business and public) \cite{rana2020impact,demirbas2020re,konig2020adapting}. National lockdowns have forced organizations to reorganize daily work in remote and online modes practically overnight \cite{rana2020impact,iivari2020digital} and digital transformation has further accelerated \cite{agostino2021new,priyono2020identifying,harmand2021digitalisation}). This situation has emphasized the crucial issue of evaluating human adaptation to digital changes and potential factors that can affect employee adaptation \cite{trenerry2021preparing,makowska2021evaluating}, such as self-efficacy in coping with ICT hassles, for example, issues with the internet connection \cite{day2012perceived,leonardi2021covid,makowska2021evaluating}. This situation highlighted the importance of research in the field of digital transformation stress \cite{makowska2021evaluating,makowska2023measuring} and the identification of the main DTS factors. 
Then we find an efficient way to measure both. Consequently, the development action plan to counteract DTS effectively. The increase in the overall level of stress in the workplace is related to the health problems reported by employees \cite{bakker2014burnout,day2012perceived}. Similarly, an increase in DTS can also directly lead to burnout and / or health problems. Therefore, during the DT project, DTS should be regularly monitored among employees. However, time-consuming psychometric surveys \cite{fagarasanu2002measurement} can be ineffective, due to the high dropout rate \cite{smoktunowicz2021resource,makowska2021evaluating} and additional disruptors for employees. Therefore, sentiment analysis (SA) may be a more efficient way of detecting symptoms of DTS \cite{makowska2021evaluating} and, for example, the ICT helplessness of employees.

\subsection{Gender and Digital Transformation Stress}
Gender stereotypes are generalizations about the attributes, abilities, and skills of men and women \cite{gupta2009role}. Because gender stereotypes are widely shared and automatically activated, they are very impactful, especially in domains that are perceived as male gender-typed \cite{cai2017gender}, especially science, math, technology and engineering (STEM) occupations \cite{haines2016times}. For example, Smith, Morgan and White\cite{smith2005investigating} showed that women are stereotypically perceived as worse at computer technology (CT) \cite{vimalkumar2021contextualizing}. These gender stereotypes are also present in the diverse employment settings and have important consequences for employee stress. Research has reported that women are facing unique stressors, with discrimination and stereotyping among others \cite{geller1994gender,liu2008use}. 
But even when the workplace setting is free from discrimination, stereotypes can shape the experience of employees through the threat mechanism of stereotypes. Stereotype threat is defined as the worry that one may confirm negative stereotypes about one's group by performing difficult tasks in the domain relevant to stereotypes \cite{steele1995stereotype}. Research on stereotype threat has shown that stereotype threat determined the perception of technical difficulties by women during completion of ICT tasks and may lead to greater stress \cite{koch2008women}. In the workplace, this psychological pressure can lead to negative emotions and increased stress \cite{roberson2007stereotype}. Therefore, we assume that, in the context of rapid digital transformation, it may lead to higher levels of digital transformation stress.

\begin{figure}
    \centering
    \includegraphics[width=0.9\linewidth]{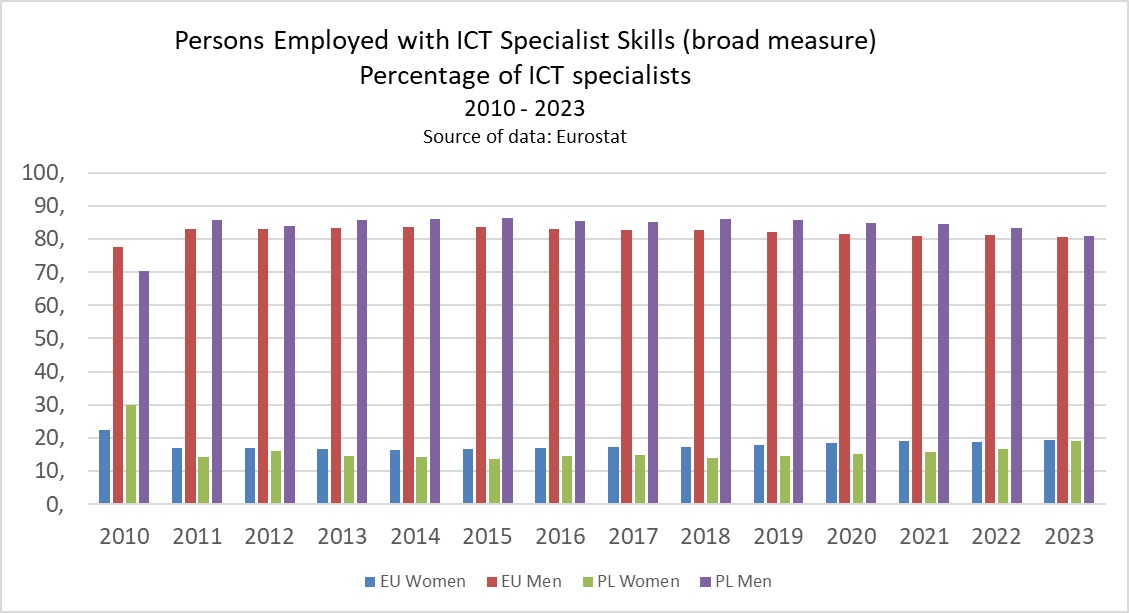}
    \caption{Share of women vs men employed in IT in EU and Poland}
    \label{fig:enter-label}
\end{figure}


\begin{figure}
    \centering
    \includegraphics[width=0.9\linewidth]{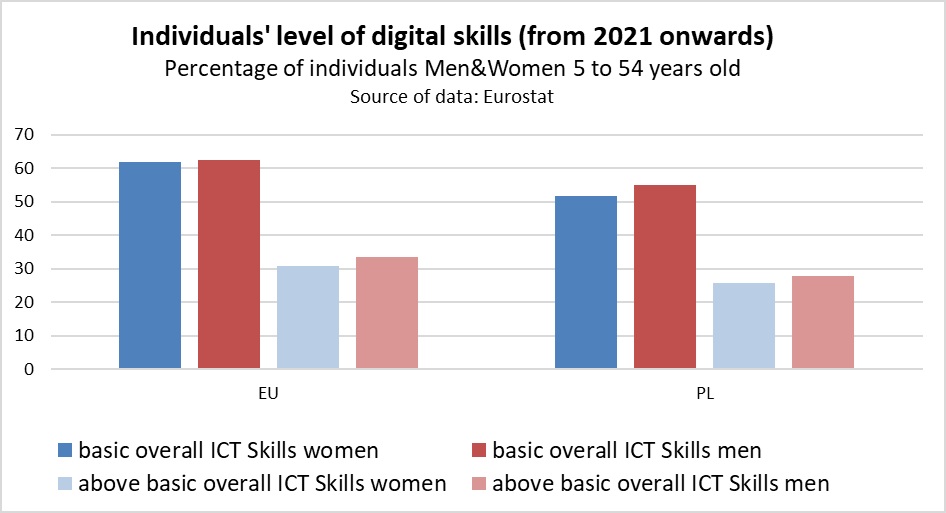}
    \caption{Level of digital skills of men and women in EU and Poland}
    \label{fig:enter-label}
\end{figure}

\subsection{Sentiment Analysis of Help Desk Tickets}

The content of Help Desk tickets written by employees to receive support in ICT-related issues is a promising source of behavioral data related to employee digital transformation stress and ICT helplessness. 
We defined ICT helplessness as the declared inability to deal with the issue in solving ICT problems. Based on similar studies \cite{makowska2021evaluating,zainuddin2014sentiment}, we use sentiment analysis to extract both negative emotions and ICT helplessness present in these short messages.
For our analysis, we collected all tickets registered between 2019 and 2020. We selected two years to examine if there is a difference in the HD sentiment analysis before and during the COVID-19 pandemic, as well as before and during important organizational changes in the company selected for this investigation. At the beginning of 2020, the Board of Directors was changed, followed by the introduction of a new organizational structure and the initiation of new DT projects. All of the above resulted in increased challenges for the employees.
In total, we collected 17515 ticket records (years 2019-2020). We started with an analysis comparing the number of HD tickets related to ICT issues between 2019 and 2020 in total, and HD tickets divided into those submitted by women and men. The help desk application was implemented in the company in first months of 2019, so the texts of previous tickets were not available in the SQL database to build the SA algorithm.

\section{Method}

\subsection{Study Assumptions}

To examine the difference between female and male employees regarding the DTS and sentiments analysis of negative emotions and ICT helplessness, we assumed that: (1) the organisation has been under DT process (2) employees have to be affected by DTS; (3) we had access to employees’ written official communication, like help desk tickets, to conduct sentiment analysis. Then we had the opportunity to examine the relationship between the results of the DTS and HD sentiment analysis, as well as to identify ICT helplessness and measure the differences between women and men in this area.

\subsection{Procedure and Participants}

The research was carried out in one of the multinational financial companies which, since 2017, has been undergoing a digital transformation project that touches on all areas of its activity. The main objective of this DT project has been to digitize the business processes in the company, among others, by the implementation of the Document Management System (DMS). The company where the research was conducted was, at the time, a medium-sized company with a permanent workforce of about 160 full-time employees and about 40 associates. The majority of the employees (78\%) had a university degree, each of the employees used IT tools, either an ERP system or a popular system used for office purposes (word processor, spreadsheet) and were also covered by the EOD implementation. Most of the employees had been in the workforce for more than 10 years. 
The company's employees who actively participated in the investigation comprised the sample of 37 participants who provided valid responses (53\% of all registered) to the DTS survey \cite{makowska2021evaluating} . Of those who completed the survey in full, 30 women and 7 men, mean age = 40.5, \textit{SD} = 8.76. 26 of the women surveyed had university education (92 \%) while of the men surveyed 3 (near 50\%) had a university degree and 3 had a high school degree (near 50\%). The survey included employees who had managerial (9 women, 2 men), expert (12 women, and 3 men) and operational (5 women and 2 men) positions. 
We have examined the sentiment in help desk tickets \cite{makowska2021evaluating}, which included 266 users (144 women and 122 men).  In fact, after matching the logins of HD users and survey participants, the survey sample consisted of 14\% of all HD users. 
All data were analyzed to acquire the number of logins for each record, that is, waw080, waw079, to map the association between the survey participants and HD users to examine gender differences in the DTS indicator. Then we analyzed negative sentiment and ICT helplessness in the HD tickets reported. All data were anonymized.

\subsection{Ethics and Company Consent}

The research protocol was approved by the Ethics Committee of our University. The present study was conducted according to the ethical standards adopted by the American Psychological Association (APA, 2010). Before participation, all participants were informed about the general purpose of the investigation and the fact that their data will remain anonymous. After marking informed consent to the study, the questionnaire was activated. Participation was voluntary and participants did not receive compensation for their participation in the study.
The company's management agreed to conduct a study of a psychometric survey and a sentiment analysis on requested written service tickets on the company's help desk system.

\section{Results}

\subsection{Sentiment Analysis Related to ICT Helplessness}

Based on the previous study on sentiment analysis of help desk tickets, we started the analysis of helplessness markers in HD tickets with specific, basic words phrases where users declared their lack of self-efficacy (Lloyd, Bond and Flaxman, 2017) in dealing with ICT hassles (e.g., software and hardware issues, connections issues, computer freezing etc.).We grouped the ICT helplessness expressions and words syntax like:
\begin{enumerate}
    \item  admitting lack of skills: “\textit{I don’t know}”, “\textit{I am not able}”,\textit{ “I don’t understand why}”, “\textit{I’m not familiar with”, “I don’t understand”, “I have forgotten how}” etc.;
    \item admitting inability to dealing with, like: “\textit{I’m not able to solve}”, “\textit{I can’t do}”
    \item directly asking for help like: “\textit{help!”, “I need your help with}”, “\textit{please, help me}”
    \item personalising the system like: “\textit{my computer yells at me}”, “\textit{system behaves strange}”
\end{enumerate}
Next, we analysed the frequency of occurrence of word syntax in HD tickets in all HD categories and categories related to ICT issues. We observed that both the number of HD tickets and the ICT helplessness markers were higher in 2020 and this number was much higher among women than among men. In 2019 there were 3867 HD tickets in total, 3047 HD tickets of ICT categories. Women reported ICT issues 79,28\% more often than men (1956 female ICT HD tickets and 1091 male ICT HD tickets). In 2020 we observed a dramatic increase of 350\% in the total number of registered HD tickets (13648), and women registered 63\% (8381) more HD tickets than men (5267). We observed that in 2020, despite the higher number of HD tickets reported by women, the proportion between the number of HD tickets reported by men and women was lower than in 2019. It shows that the 2020 as COVID-19 could affect both men and women in relation to ICT hassles.

\subsection{Gender Differences in Digital Transformation Stress Among Employees }

We used all 2020 HD tickets with negative emotion markers for sentiment analysis. There was a significant and positive correlation between the digital transformation stress score (DTSS) \cite{makowska2023measuring} and the results of the HD sentiment analysis, as well as a significant and positive correlation between Perceived Stress Scale (PPS-4) \cite{cohen1983medida} and HD sentiment analysis results.
Although the sample of users who completed the psychometric survey was rather small (28 matched users), we decided to examine if there is a difference between male and female employees in the level of experienced DTS. Analyses using Student’s t test for independent samples revealed that there was a significant difference in the DTS level between male and female employees, \textit{t} (25) = 2.70; \textit{p} = .012, and female employees reported a higher DTS level  (\textit{M} = 3.01,\textit{ SD} = 0.06) than male employees (\textit{M} = 2.21, \textit{SD} = 0.47). However, there was no significant difference in general stress (PSS-4) level between male and female employees, \textit{t} (26) = 1.75; \textit{p} = .092.

\subsection{Gender Differences in Help Desk Tickets Analysis}

Similar gender differences were observed in the number of HD tickets registered in 2019 and 2020, as well as the number of tickets with negative emotion markers or with helplessness markers. In 2019 female employees reported on average 13.58 (\textit{SD} = 21.98) HD tickets of ICT categories compared to 8.93 \textit{ (SD} = 12.74) tickets reported by men and this difference was significant \textit{t} (264) = 2,06; \textit{p} = .040. Among these tickets, there were also differences in negative sentiment level, with women reporting a higher number of ICT HD tickets with markers of negative sentiment (\textit{M} = 7.15, \textit{SD} = 11.57) than male workers (\textit{M} = 4.14, \textit{SD} = 6.59) t (264) = 2.54, \textit{p} = .012. We observed a consistent pattern in HD ICT tickets with ICT helplessness markers – female employees reported a higher number of such HD tickets (\textit{M} = 5.56,\textit{ SD} = 15.27) than male employees (\textit{M} = 2.61,\textit{ SD} = 5.21), \textit{t}  (264) = 2.04,\textit{ p} = .042. 

\begin{figure}
    \centering
    \includegraphics[width=0.9\linewidth]{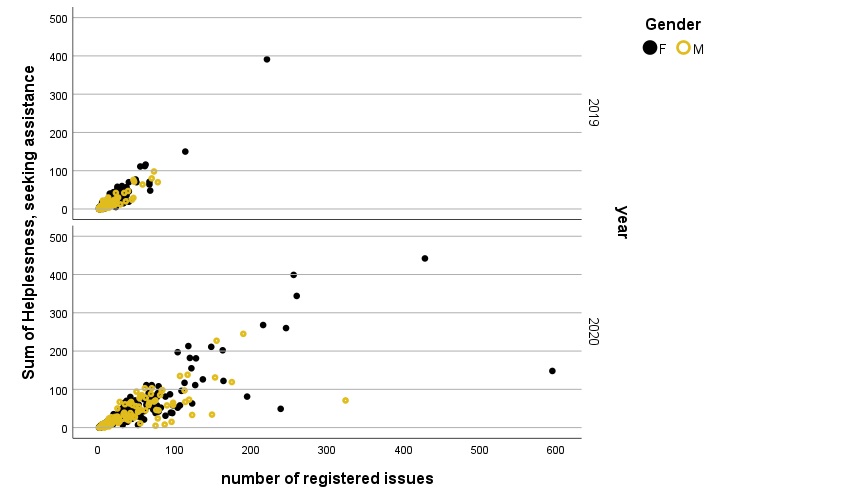}
    \caption{The relationship between the registered issues and the intensity of the negative sentiment markers.}
    \label{fig:enter-label}
\end{figure}

\begin{figure}
    \centering
    \includegraphics[width=0.9\linewidth]{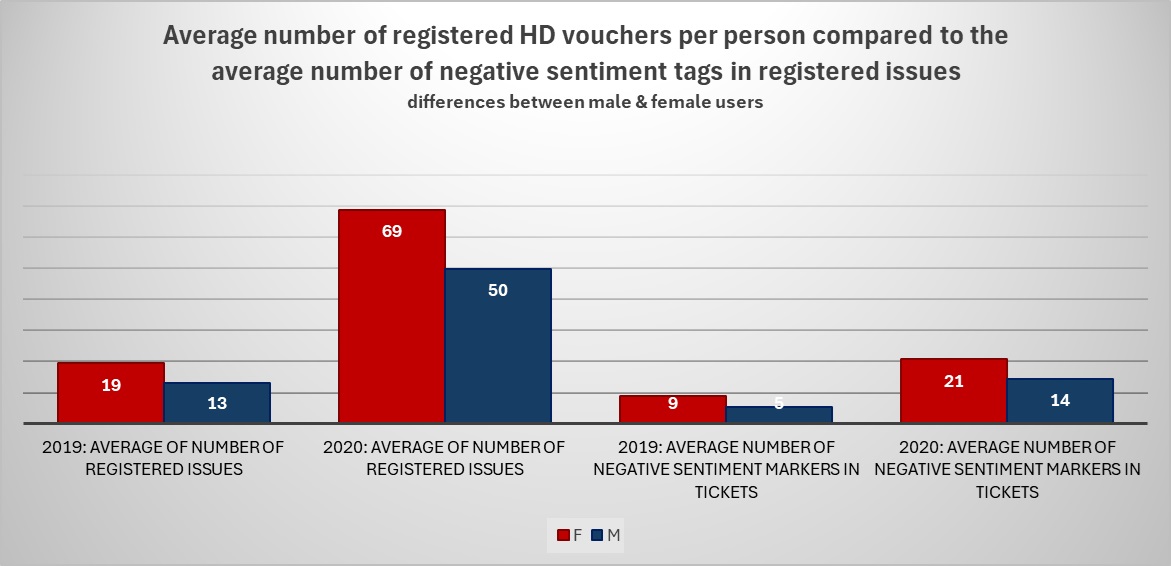}
    \caption{Average tickets and negative sentiment tags per person.}
    \label{fig:enter-label}
\end{figure}

The same aspects were analyzed in 2020, during the COVID-19 pandemic, and generally the employees reported a similar number of HD ICT tickets with markers of negative sentiment and helplessness as in 2019. In addition, the same pattern of gender differences was observed. Again, a higher number of ICT HD tickets was registered by female employees (\textit{M} =32.35, \textit{SD} =38.68) in comparison to male workers (\textit{M} = 23.65 , \textit{SD} = 25.85 , \textit{t} (264) = 2.12,\textit{ p} = .035). We revealed the higher number of tickets with negative sentiment in ICT HD tickets reported by female employees (\textit{M} = 14.48, \textit{SD} = 17.59) compared to male workers (\textit{M} = 9.84, \textit{SD} = 14.90, \textit{t} (264) = 2.30, \textit{p} = .022). Women also submitted a higher number of HD tickets with helplessness markers (\textit{M} = 9.53, \textit{SD} = 13.35) than the male workers (\textit{M} = 6.52, \textit{SD} = 11.75, \textit{t}  (264) = 1.93, \textit{p} = .052.

\section{Discussion and Future Work}

The main aim of the present study was to provide preliminary evidence that gender is an important predictor of Digital Transformation Stress evaluated two-fold – in a self-report psychological scale, namely Digital Transformation Stress Scale, and in sentiment analysis of Help Desk tickets. The results consistently indicated that female employees experience more DTS than male employees. Firstly, our results revealed that the gender of the employees differentiated the level of DTS on a self-reported scale. Female employees described their DTS as higher compared to male employees. The sentiment analysis of help desk tickets showed that female employees reported a higher number of tickets with negative sentiment and with ICT helplessness markers. Consequently, these gender differences were observed in HD tickets (both in all categories and in categories related to ICT issues) of the wave reported in 2019 and 2020. \textbf{The results can be interpreted as showing that female workers experience a higher level of digital transformation stress in the workplace.} It is also important to note that the gender gap observed in DTS is not reflected in the general level of stress at the workplace. Based on self-efficacy theory related to stress \cite{bandura1989human,preacher2008asymptotic}, especially in the context of the workplace, this gender gap in digital transformation stress can be explained in terms of lower self-efficacy of women when dealing with CT difficulties. This self-efficacy can be shaped by gender stereotypes that describe women as having lower skills and abilities in technology \cite{smith2005investigating}. Similar effects may be caused by the stereotype threat experienced by women when using technology \cite{koch2008women}. This additional worry that others will perceive you through the lens of negative gender stereotypes can cause negative emotions \cite{bedynska2015stereotype,schmader2008integrated}, the feeling of helplessness and lead to burnout \cite{bedynska2015stereotype,hall2015engineering}, and difficulties in acquiring new skills \cite{mangels2012emotion}. The latter may lead to lower ICT literacy among women.

These results, although they may seem theoretical, have real impact on IT tools' design and company policies in regarding ICT tools, procedures, and support systems. First, if gender affects the way women view and communicate their accomplishments, should AI tools designed to evaluate employees and their performance be adjusted to take this into account? Additionally, there are some lessons to be learned for the design of in-company IT-systems, such as ticketing. Our results show that women use the "normal" issue priority more often than men, who prefer high priority. Men also, on average, use more words to describe the problems they encounter. This may be due to worries about being perceived through the lens of stereotypes around women and technology.  As women experience higher DTS it may also be important to implement automatic stress-detection systems based on written office communication and to offer appropriate interventions to lower this stress and increase support, to prevent decrease in well-being, burnout and employee turnover.
It is worth mentioning here that, of those who took part in the DTS survey, the majority of women volunteered to participate in participatory workshops to develop online interventions to cope with the stress of digital transformation. As a result, the workshops held afterwards were attended by women.

\subsection{Limitations}

The present study has several limitations. Since 2021, we have not had more access to the Help Desk ticket data, nor have we had the ability to conduct the psychometric survey in the same organization to match employee logins with HD system logins. Moreover, the syntax of the Polish language makes the development of an algorithm for sentiment analysis a challenge. However, our results encourage further work in this area.

\section{Conclusions}

Our research confirmed that women experience higher levels of DTS and ICT helplessness at work, which is likely due to stereotypes that affect women during the digital transformation process. Consequently, addressing these factors can be an important element that contributes to improving employee well-being. Moreover, our interdisciplinary research confirmed that using alternative methods of detecting stress symptoms, namely sentiment analysis of help desk tickets instead of time-consuming surveys, is efficient at identifying DTS symptoms, e.g., ICT helplessness. This research paves the way for the creation of automated stress detection systems in a workplace, which could help address employee stress with specifically tailored interventions at an early stage, before it starts to have adverse effects on employee well-being and performance.
\bibliographystyle{plain}
\bibliography{bibliography}
\end{document}